\documentclass[aps,prb,superscriptaddress,reprint,showpacs,floatfix
]{revtex4-1}

\usepackage{tabularx}
\usepackage{float}
\usepackage{bm}
\usepackage{graphicx}
\usepackage{amsmath}
\usepackage{mathtools}
\usepackage{color}
\usepackage{mathrsfs}
\usepackage[colorlinks=true,citecolor=blue]{hyperref}
\hypersetup{colorlinks=true,citecolor=blue,linkcolor=blue,urlcolor=blue}

\usepackage{chngcntr}

\begin{document}

\title{All optical resonant magnetization switching in $\text{CrI}_3$ monolayers}

\author{A. Kudlis}
\affiliation{Department of Physics and Engineering, ITMO University, St. Petersburg, 197101, Russia}
\email{andrew.kudlis@metalab.ifmo.ru}

\author{I. Iorsh}
\affiliation{Department of Physics and Engineering, ITMO University, St. Petersburg, 197101, Russia}
\email{i.iorsh@metalab.ifmo.ru}

\author{I. A. Shelykh}
\affiliation{Department of Physics and Engineering, ITMO University, St. Petersburg, 197101, Russia}
\affiliation{Science Institute, University of Iceland, Dunhagi-3, IS-107 Reykjavik, Iceland}
\email{shelykh@hi.is}

\begin{abstract}
Efficient control of a magnetization without an application of the external magnetic fields is the ultimate goal of spintronics. We demonstrate, that in monolayers of $\text{CrI}_3$, magnetization can be switched all optically, by application of the resonant pulses of circularly polarized light. This happens because of the efficient coupling of the lattice magnetization with bright excitonic transition. $\text{CrI}_3$ is thus perspective functional material with high potential for applications in the domains of spintronics and ultra-fast magnetic memory.
\end{abstract}

\maketitle

\textit{Introduction.}
The needs of information processing demand performing of low cost, high speed, and high-density magnetic recording, which does not need the application of external magnetic fields. Achieving of this aim is among the main goals of spintronics. In conventional semiconductors, the spin of electrons can be controlled by application of the electric field via Rashba spin orbit interaction \cite{DattaDas1990,Ganichev2004,Koo2009}. However, magnetic devices based on electrostatic control of individual spins have certain practical disadvantages, related to the necessity of reaching of sub Kelvin temperatures and limitations for characteristic times of the spin inversion. Therefore, the search of novel magnetic materials and devices which reveal efficient and controllable magnetization switching continues. 

The possibility of an optical control of magnetization is of special interest here, as it can potentially push the speed of the magnetic reversal and associated magnetic memory writing speed towards THz frequencies. Optical magnetization switching was very recently demonstrated for CdFeCo \cite{Ignatyeva2019,Alives2020,Stanciu2020,Davies2020,Igrashi2020} and TbFeCo \cite{Lu2018} ferromagnetic alloys, as well as in Co/Gd bilayers \cite{vanHees2020}. 

In this contect, the family of functional 2D materials, namely chromium dichalcogenides, such as CrI$_3$ and CrBr$_3$, is of special interest, as they posses a unique combination of optical and magnetic properties. In particular, they demonstrate robust optical excitonic response, with record high values of excitonic binding energies and oscillator strengths \cite{Wu2019}, exceeding even the values reported for transition metal dichalcogenides \cite{Chernikov2014,Splendiani2010,Steinleitner2017,Wang2018}. In the same time, these materials are 2D Ising ferromagnets  \cite{Huang2017,Zheng2018,Kashin2020}, thus having an additional twist, related to the giant Zeeman splitting of the valence and conduction bands, which, among the rest, strongly affects their optical properties, leading to such phenomena as giant Kerr response\cite{Huang2017}, magnetic circular dichroism \cite{Seyler2018} and onset of 2D magnetoplasmons \cite{Pervishko2020}.

\begin{figure}[t!]
\centering
\includegraphics[width=0.99\columnwidth]{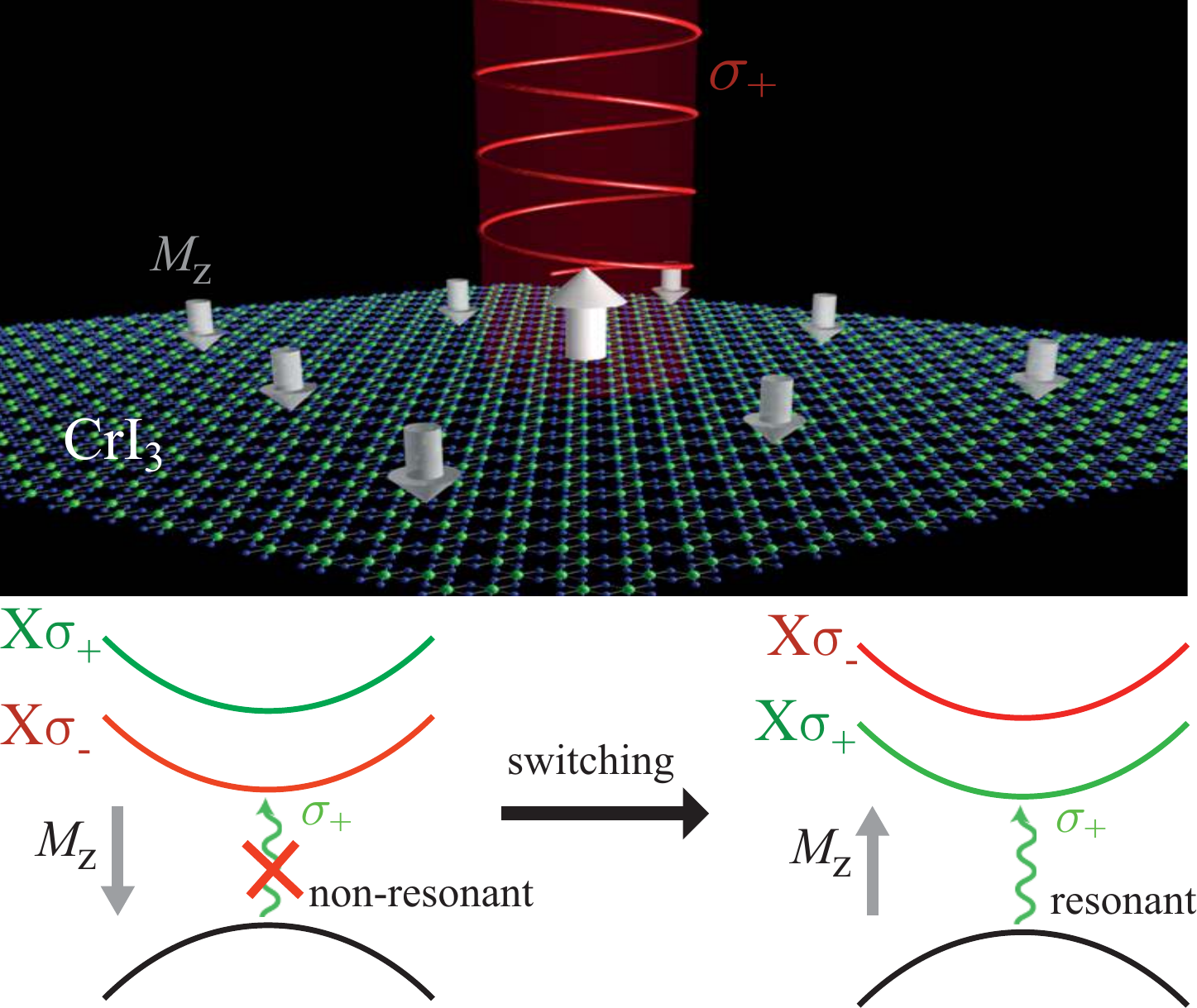}
\caption{(upper panel) A monolayer of CrI3 is irradiated by a circularly polarized light which reverses its magnetization. (lower panel) The mechanism of the magnetization switching. Circular polarized pump is not resonant with corresponding exciton transition if magnetization is pointing down (left part), but becomes resonant if the direction of the magnetization is switched.}
\label{Fig_geom}
\end{figure}

In the present paper we demonstrate, that the combination of the pronounced excitonic and fertomagnetic responses leads to the possibility of polarization selective switching of the magnetization. The main idea of the proposed effect is illustrated schematically in the Fig.~\ref{Fig_geom}. The band structure of a monolayer of CrI$_3$ is shown schematically in~Fig.~\ref{Fig_geom}(lower panel). As CrI$_3$ is a direct band semiconductor, optical transition is allowed from the top valence band to the bottom of the conduction band. Attraction between an electron and a hole leads to the formation of strongly coupled bright excitons. Depending on the direction of the magnetization of a sample, ground state excitons can be excited by $\sigma^+$ or $\sigma^-$ light. 

Consider the geometry shown in the lower panel of Fig.~\ref{Fig_geom}, when ground state corresponds to $\sigma^-$ excitation, and we send a pulse of $\sigma^+$ light in resonance with it. As $\sigma^+$ transition is strongly offset in energy due to the giant Zeeman splitting produced by the lattice magnetization, absorption of $\sigma^+$ photons is only virtual. However, this virtual absorption creates an effective magnetic field, which tends to invert the magnetization. This process is favoured by the fact, that magnetization switch will make optical absorption resonant, and if the intensity of the optical pump exceeds some threshold value, it finally happens.

\textit{Methods.}
To describe the process of magnetization switching, we proposed a phenomenological model based on the system of coupled driven-dissipative Gross-Pitaevskii type equations for the concentration of spin polarized excitons with Landau-Lifshiz-Gilbert equation for the lattice magnetization. 

The total effective magnetic field, affecting the dynamics of the magnetization can be estimated as the sum of a real magnetic field $\boldsymbol{H}_0$, magnetic field provided by spin-spin interactions $\boldsymbol{H}_{ex}$  and an effective provided by spin polarized excitons:
\begin{equation}
    \boldsymbol{H}=\boldsymbol{H}_0+\boldsymbol{H}_{ex}+\delta \left(|\psi_{+}|^2-|\psi_{-}|^2\right)\boldsymbol{e}_z,\label{effective_field}
\end{equation}
where $\psi_{+}$ and $\psi_{-}$ are order parameters which correspond to the densities of excitons with spin up and spin down respectively, $n_\pm=|\psi_{\pm}|^2$, and 
\begin{equation}
\boldsymbol{H}_{ex}=A\,\nabla^2\boldsymbol{M}+D\,\nabla\times\boldsymbol{M},    \label{Hex}
\end{equation}
where the constants $A$ and $D$ characterize the exchange interaction of Heisenberg and Dzyaloshinskii-Moriya type respectively \cite{Albert2019}. 

This effective field enters into the Landau-Lifshiz equation describing the magnetization dynamics~\cite{Albert2019,leliaert2019tomorrow}:
\begin{eqnarray}
&&\dfrac{\partial\boldsymbol{M}}{\partial t}=-\gamma[\boldsymbol{M}\times \boldsymbol{H}] -\dfrac{\gamma\eta}{M_s}[\boldsymbol{M}\times[\boldsymbol{M}\times \boldsymbol{H}]], \ \
\end{eqnarray}
where $\boldsymbol{M}$ is magnetization, $M_s$ is its saturation value, $\gamma=\mu_0\gamma_0$, $\gamma_0=g\mu_B/\hbar$, and $\eta$ is dimensionless damping constant.
To close the system of the equations, one also needs to write the equations for the dynamics of the excitonic fields, which in the simplest case can be chosen in the form of the driven-dissipative Gross-Pitaevskii equations for the  components $\psi_{+}$ and $\psi_{-}$ \cite{High2013,Vishnevsky2013}:
\begin{eqnarray}
&&i\hbar\dfrac{\partial\psi_{+}}{\partial t}=-\frac{\hbar^2}{2m_X}\nabla^2\psi_{+}+\beta M_z\psi_{+} \nonumber\\
&&\qquad\qquad\qquad\qquad\qquad\qquad+\alpha|\psi_{+}|^2\psi_{+}+P_{+}e^{i\delta_{+}t},\\
&&i\hbar\dfrac{\partial\psi_{-}}{\partial t}=-\frac{\hbar^2}{2m_X}\nabla^2\psi_{+}-\beta M_z\psi_{-} \nonumber\\
&&\qquad\qquad\qquad\qquad\qquad\qquad+\alpha|\psi_{-}|^2\psi_{-}+P_{-}e^{i\delta_{-}t}\label{phi_down},
\end{eqnarray}
where $m_X$ is excitons mass, $P_{+}$ and $P_{-}$ are pump amplitudes in the right and left circular polarized components respectively.
They are related with corresponding pump power densities $W_\pm$ as follows: $W_{\pm}=P^2_{\pm}\omega_{ex}/\hbar f_0$, where $\omega_{ex}$ is exciton frequency and $f_0$ is its oscillator strength. The coupling constant $\delta$ and $\beta$ are  related to each other, as both of them describe the mutual action of the excitonic and magnetic subsystems. To estimate the corresponding relation, one can assume, that the maximum of the product $\beta M_s$, corresponding to the Zeeman splitting of the excitons should be approximately equal to the characteristic energy $\hbar\gamma H$ entering into Landau-Lifshiz equation for the case, when the concentration of polarized excitons is around one per unit cell of size $d\times d$: 
\begin{equation}
\hbar\gamma\delta/d^2\approx\beta M_s.\label{delta_beta}
\end{equation}

In our analyzis, we consider the case of the spatially homogeneous optical pump and suggest, that both $\boldsymbol{M}$ and $\psi_\pm$ does not have any spatial dependence. This will allow us to drop the term related to the exchange interaction in the expression for effective magnetic field (Eq.\ref{Hex}). Moreover, we can introduce the following set of dimensionless variables:
\begin{eqnarray}
    &\tilde{\boldsymbol{M}}=\dfrac{\boldsymbol{M}}{M_s}, \quad \tilde{\psi}_{\pm}=\dfrac{\psi_{\pm}}{\psi_s},\quad \tilde{t}=\dfrac{t}{t_s},\quad \tilde{P}_{\pm}=\dfrac{P_{\pm}}{P_s}&\nonumber\\
    &\tilde{W}_{\pm}=\dfrac{W_{\pm}}{W_s}, \quad \tilde{\boldsymbol{H}}_0=\dfrac{\boldsymbol{H}_0}{M_s}, \quad \tilde{\boldsymbol{H}}=\dfrac{\boldsymbol{H}}{M_s}\nonumber&
\end{eqnarray}
This will allow us to rewrite the system of the dynamic equations in the following form: 
\begin{eqnarray}
    &\tilde{\boldsymbol{H}}=\tilde{\boldsymbol{H}}_0+\left(|\tilde{\psi}_{+}|^2-|\tilde{\psi}_{-}|^2\right)\boldsymbol{e}_z\label{effective_field_dl},&\\
    &\dfrac{d\tilde{\boldsymbol{M}}}{d\tilde{t}}=-[\tilde{\boldsymbol{M}}\times \tilde{\boldsymbol{H}}] -\eta[\tilde{\boldsymbol{M}}\times[\tilde{\boldsymbol{M}}\times \tilde{\boldsymbol{H}}]],& \\
    &i\dfrac{d\tilde{\psi}_{+}}{d\tilde{t}}=a_1 \tilde{M_z}\tilde{\psi}_{+} +a_2|\tilde{\psi}_{+}|^2\tilde{\psi}_{+}+\tilde{P}_{+}e^{i \tilde{\delta}_{+}\tilde{t}},&\\
    &i\dfrac{d\tilde{\psi}_{-}}{d\tilde{t}}=-a_1\tilde{M_z}\tilde{\psi}_{-} +a_2|\tilde{\psi}_{-}|^2\tilde{\psi}_{-}+\tilde{P}_{-}e^{i \tilde{\delta}_{-}\tilde{t}}\label{phi_down_dl}.&
\end{eqnarray}
where the corresponding dimensionless parameters expressed in terms of the original ones are given in Table~\ref{tab:par}. 
\begin{table}[b!]
 \centering
    \caption{Definitions of parameters which enter the system of eqs.~\eqref{effective_field_dl}-\eqref{phi_down_dl}. The quantities designed to make the variables dimensionless are also present in the table. The corresponding numerical values are calculated on the basis of numbers for $M_s$, $\delta$, $d$, $\omega_{ex}$, $f_0$, and $\alpha$ discussed within the body of the paper} %
    \label{tab:par}
    \begin{tabular}{cll|cll}
      \hline
      Par. & Definition & Value &Par. & Definition& Value  \\
      \hline
  $a_1$& $\delta/d^2\,M_s$ &$2134$ & $t_s$& $1/\gamma\,M_s$ &$2.33\cdot10^{-2}\,$ns     \\
  $a_2$& $\alpha/\delta\gamma\hbar$ & $1.663$&      $\psi_s$& $\sqrt{M_s/\delta}$ & $2.16\cdot10^5\,$cm$^{-1}$\\ 
  $\tilde{\delta}_{+}$& $\delta_{\pm} \, t_s$ & $11.04$ & $P_s$& $\psi_s\hbar/t_s$ &    $9.77\cdot10^{-17}\,$J$\cdot$cm$^{-1}$  \\
  $\tilde{\delta}_{-}$& $\delta_{\pm} \, t_s$ & $0.000$ & $W_s$& $P_s^2\omega_{ex}/\hbar\,f_0$ &$4.12\cdot10^{-1}\,$W$\cdot$cm$^{-2}$\\
    \hline
    \end{tabular}
\end{table}
\begin{figure}[t!]
\centering
\includegraphics[width=0.99\columnwidth]{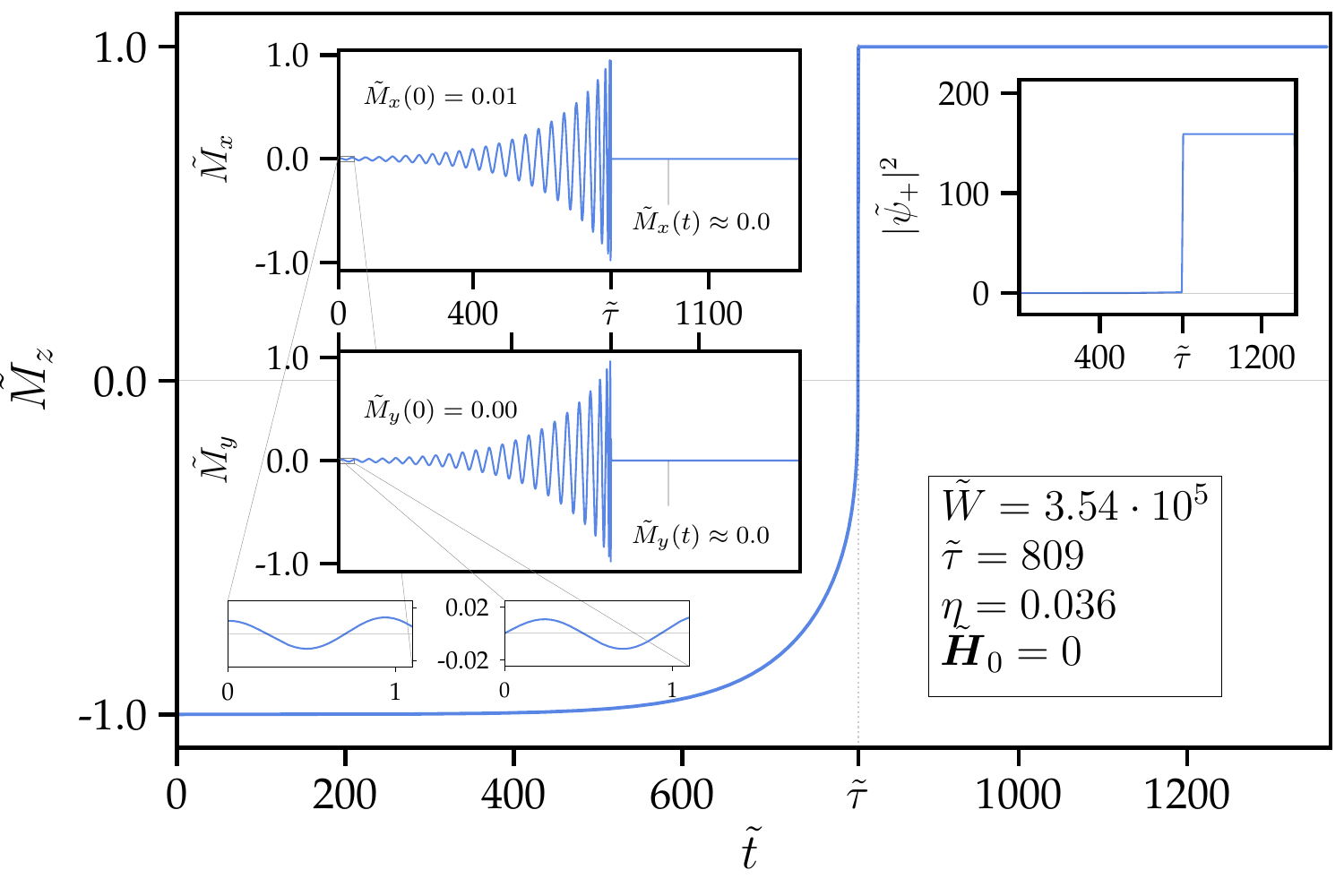}
\caption{Dependencies of magnetization components $\tilde{M}_x$, $\tilde{M}_y$, and $\tilde{M}_z$ as well as the density $|\tilde{\psi}_{+}|^2$ on time obtained by solving the equations~\eqref{effective_field_dl}-\eqref{phi_down_dl} in case the absence of the external magnetic field~($\tilde{\boldsymbol{H}}_0=0$). The transition time is denoted by $\tilde{\tau}$. The initial state corresponds to no excitons present in the system, and magnetization pointing down. After characteristic transition time the direction of the magnetization switches, and excitonic concentration increases in a step like manner. In plane components exhibit oscillations in the transient regime, but are zero after the switching.}
\label{mag_beh}
\end{figure}
As for numerical values of the parameters, we use the following numbers. The value of exciton frequency $\hbar \omega_{ex}$ is chosen as $1.5\,$eV~\cite{D0TC01322F,Huang2017}, while $\hbar f_0$ is assumed to equal to $0.033\,$eV. The value of saturation magnetization for CrI$_3$ is chosen as $M_s=0.137\,$mA/layer~\cite{Jiang2018}. The constant $\alpha$ responsible for the nonlinearity is expected to be the same, as for the excitons in TMD monolayers \cite{Shahnazaryan2017}, and is taken as $0.1\,$eV$\cdot 1\,$nm$^2$. The estimation of $\delta$ is done using the relation~\eqref{delta_beta}  with $\beta M_s=0.06\,$eV. The characteristic size of unit cell $d$ is assumed to be $1\,$nm. The determination of Gilbert damping constant is a serious computational challenge and deserves a separate consideration, however the typical values for such layer materials are expected to be measured within the following range: $\eta\in[0.01,0.1]$~\cite{Dolui2020}. We would like to stress, that although our choice of the parameters is typical for 2D materials, their exact values for the considered material can not be defined with any satisfactory precision at the current stage of the knowledge.

\textit{Results and discussion.}
Based on the equations~\eqref{effective_field_dl}-\eqref{phi_down_dl} the dynamics of the system is analyzed. The typical behaviour of  magnetization $\tilde{\boldsymbol{M}}$ and density $|\tilde{\psi}_{+}|^2$ is shown in Fig.~\ref{mag_beh}. The initial state corresponds to no excitons present in thesystem, and magnetization pointing down.  As one can see, after characteristic transition time $\tilde{\tau}$ the direction of the magnetization switches, and excitonic concentration increases in a step like manner, as condition of the resonant absorption is achieved. In the same time, in-plane components of the magnetization exhibit oscillations in the transient regime, but remain zero after the switching.

\begin{figure}[h!]
\centering
\includegraphics[width=0.49\textwidth]{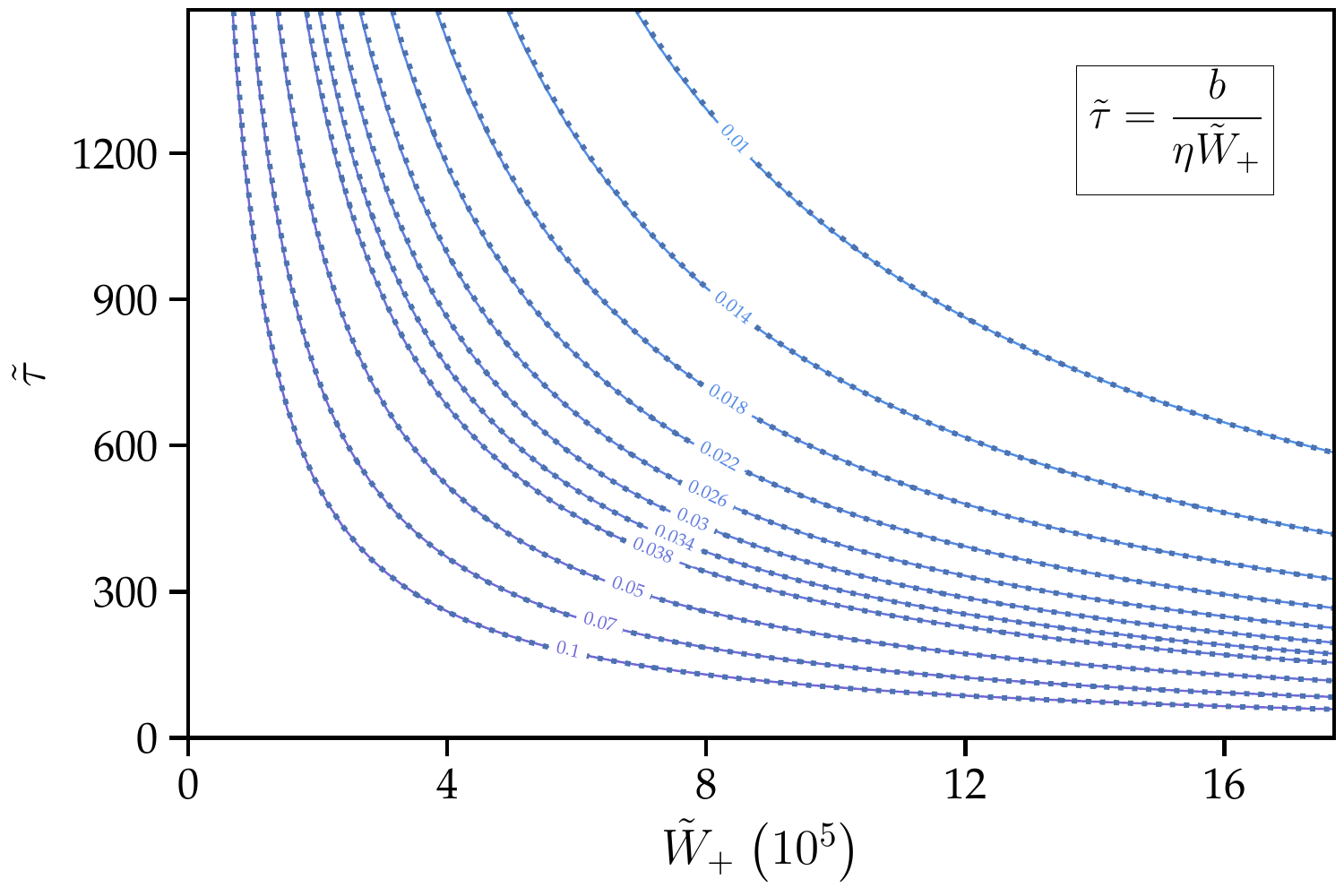}
\caption{The dependence of the transition time $\tilde{\tau}$ on power density of the circular polarized pump $\tilde{W}_{+}$ for different values of Gilbert damping constant $\eta$ for the case of the absence of the external magnetic field, $\tilde{\boldsymbol{H}}_0=0$. All the curves were fitted by the following expression~\eqref{TauFormula}(dotted lines), which gives the perfect match for $b=1.04\cdot10^7$.}
\label{tau_zero_field}
\end{figure}

The main parameter, characterizing the switching is the transition time $\tilde{\tau}$. Its dependence on the pump power density $\tilde{W}_{+}$ for the case when $\tilde{\boldsymbol{H}}_0=0$ for various values of the damping paramater $\eta$ is shown in Fig.~\ref{tau_zero_field}. Naturally, $\tilde{\tau}$ decreases with increase of $\tilde{W}_{+}$, and corresponding dependence can be perfectly fitted by the following phenomenological relation: %
\begin{equation}
\tilde{\tau}=\frac{b}{\eta\tilde{W}_{+}}. \label{TauFormula} 
\end{equation} 
For the set of the parameters corresponding to the Table 1, nanosecond switching times are reached for the pump intensities of about $10^5\,$W$\cdot$cm$^2$. Note, that in the simple model we use, where the processes of the excitonic decoherence are neglected, there is no threshold for the magnetization switching in the case when external z-directed magnetic field is absent. 

\begin{figure}[t!]
\centering
\includegraphics[width=0.99\columnwidth]{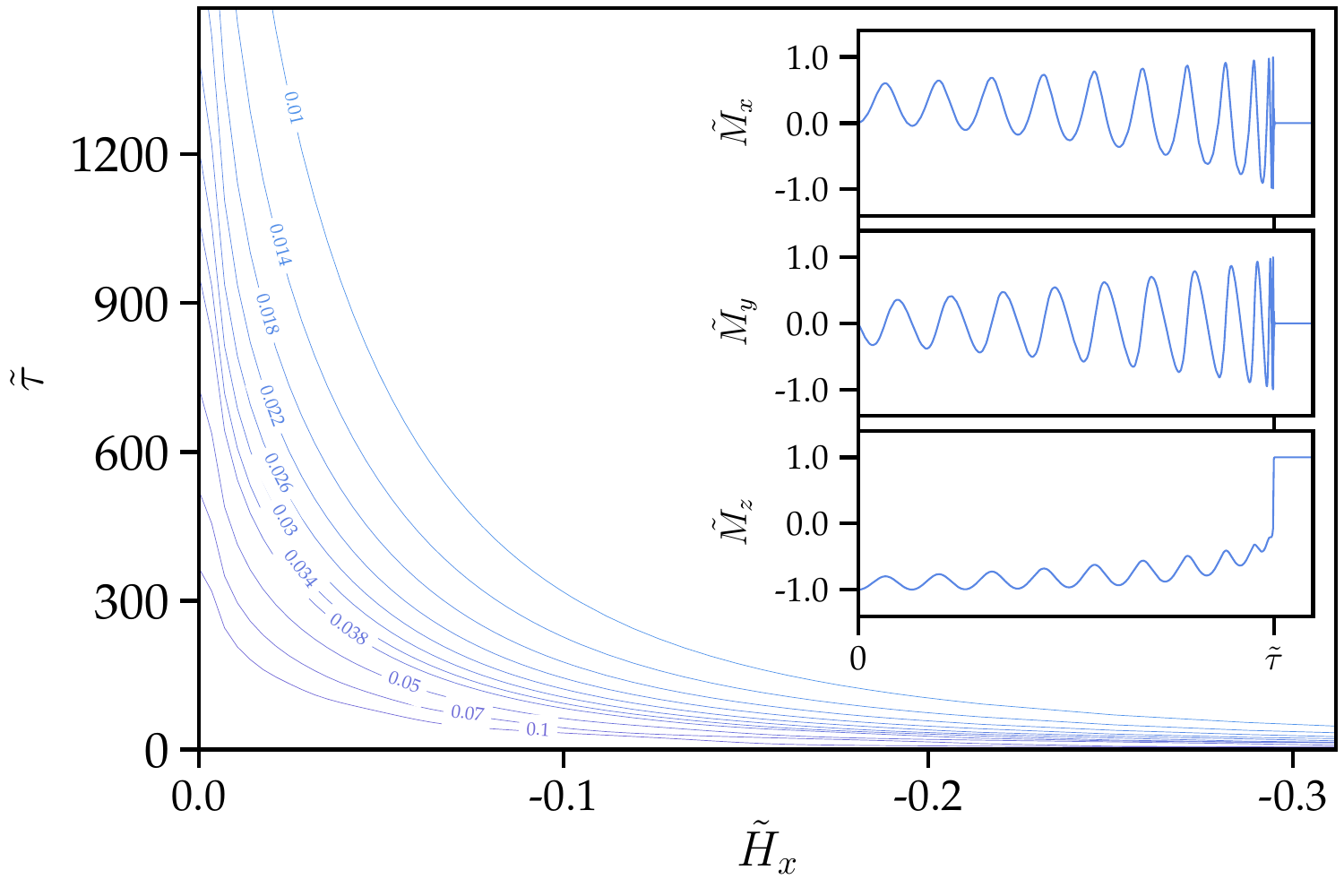}
\caption{The dependence of the transition time $\tilde{\tau}$ on the lateral external magnetic field ($\tilde{\boldsymbol{H}}_0=\tilde{H}_x\vec{e}_x$) for different values of Gilbert damping constant $\eta$ at fixed value of the intensity of circular polarized pump~($\tilde{W}_{+}=2.83 \cdot 10^5$). As one can see, lateral field favors the switching, reducing the corresponding characteristic time.}
\label{tau_lat_field}
\end{figure}

The application of the external magnetic field $\tilde{\boldsymbol{H}}_0$ strongly affects the switching process. The application of the lateral field leads to the rapid decrease of $\tilde{\tau}$ as it can be seen in Fig.~\eqref{tau_lat_field}. This effect has clear physicsl meaning, as in-plane field produces additional torque acting on z component of magnetization. In contrast to the case when $\tilde{\boldsymbol{H}}_0$ is zero, here we do not find a simple fit for $\tilde{\tau}$ on the whole field range. However, in the regime when external field are small~($\tilde{H}_x<<1$) the following approximation is valid:
\begin{equation}
    \tilde{\tau}=\dfrac{b}{\eta\tilde{W}_+(1+a\,\tilde{H}_x)^{\kappa}}, \label{TauB}
\end{equation}
where the parameters $a$ and $\kappa$ are estimated as and $120$ and $0.4$ respectively. We did not manage to get simple universal relation, describing the behavior of $\tilde{\tau}$ in the region of big strong lateral magnetic fields, but it decays faster, then in the expression \eqref{TauB}. For the choice of the parameters we use, the lateral field of $1\,$T will lead to the decrease of the switching time by $10^3$, which will allow to reduce the values of the pump needed to reach the nanosecond switching times by the same factor.

\begin{figure}[t!]
\centering
\includegraphics[width=0.99\columnwidth]{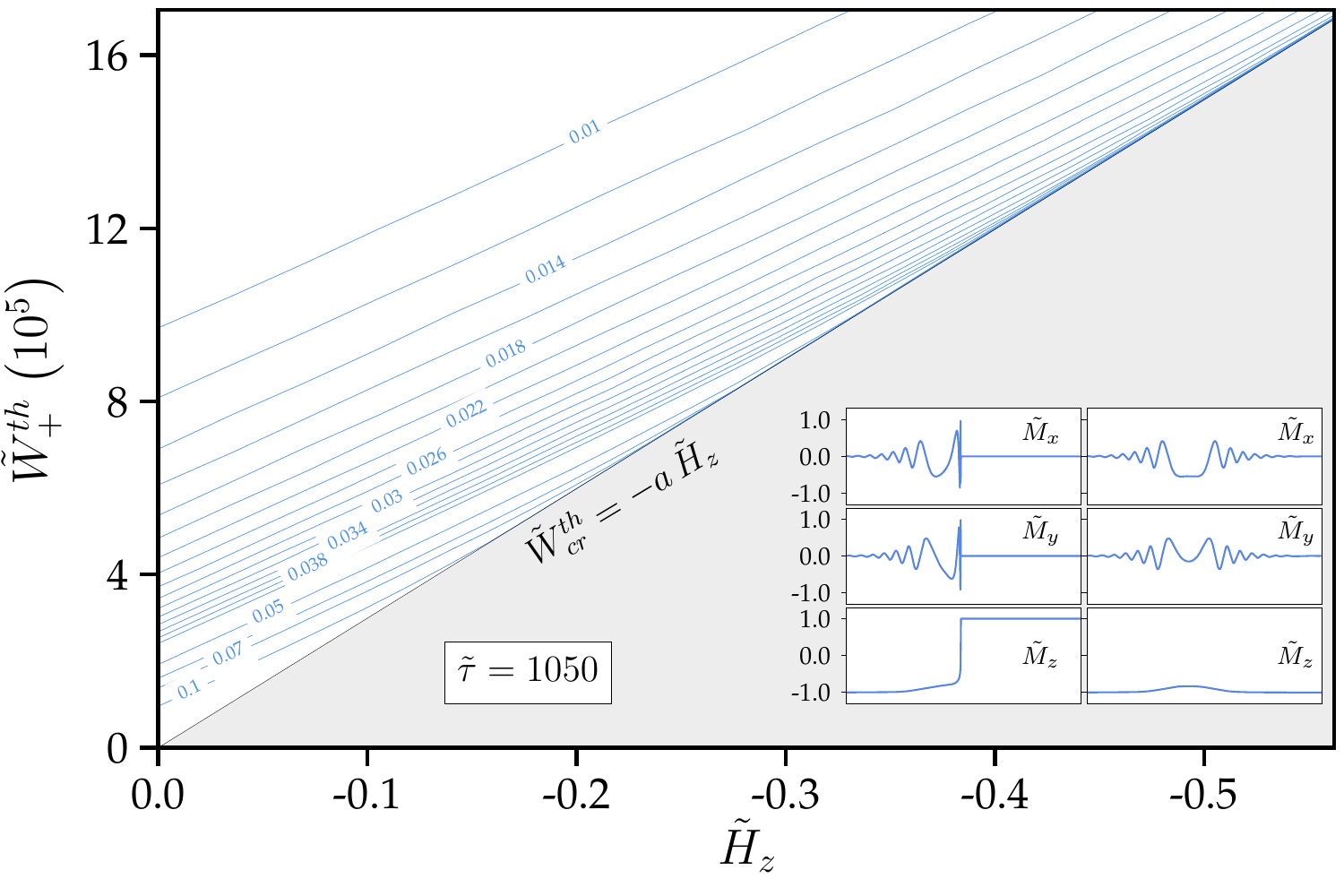}
\caption{Effect of the application of the external z-directed magnetic field $\tilde{\boldsymbol{H}}_0=\boldsymbol{e}_z\tilde{H}_z$ on polarization switching. Gray region corresponds to the regime, when switching does not occur at any time. The threshold intensity is well described Dependencies of threshold value $\tilde{W}_{+}^{th}=aH_z$. In the white region, the blue curves correspond to constant values of the switching time $\tilde{\tau}=1050$ for different values of $\eta$. It is seen, that bigger intensities are needed to keep the switching time constant with increase og $H_z$ and therefore the perpendicular magnetic field is not favorable for the switching. The insets show the temporal dependence of the components of the magnetization switching, if the intensity of the pump is above the threshold (panels A), and below it (panels B) for the case $\eta=0.1$.}
\label{w_th_hz}
\end{figure}

The case, when external magnetic field is applied along z-axis, is illustrated by Figs.\ref{w_th_hz}. In contrast to the cases of absent and lateral external fields, here the threshold for the magnetization switching appears. The threshold intensity linearly increases with increase of $\tilde{H}_z$, and if the pump is below the threshold, the switching does not occur at any time (gray region in the plot). Above the threshold, the increase of the z-directed external field leads to the decrease of the switching time, as it can be seen from the blue curves, corresponding to constant values of $\tilde{\tau}$. Thus in general, contrary to the case of the lateral field, z-directed magnetic field is not favorable for the switching.

\textit{Conclusions.}
In conclusion, we developed phenomenological theory of all optical resonant magnetization control in CrI$_3$ monolayers. It was demonstrated, that the presence of the robust bright excitonic resonances coupled to lattice magnetization leads to the possibility of polarization sensitive magnetization switching in the nonlinear regime. We investigated the dependence of the switching time on pump intensity and external magnetic field, demonstrating that lateral fields are favor the switching, reducing the switching time, while perpendicular fields have opposite effect. Our results can be used for practical development of ultra fast magnetic memory elements.

\textit{Acknowledgement.}
The work was funded by RFBR and DFG, project number  21-52-12038. IAS acknowledges support from Icelandic Research Fund (project "Hybrid polaritonics").

\bibliography{lit}

\end{document}